\documentclass[reprint,aps,prb]{revtex4-2}
\usepackage{amsmath}
\usepackage{amsfonts}
\usepackage{amssymb}
\usepackage{graphicx}
\usepackage{bm}
\usepackage{bbold}
\usepackage{hyperref}
\usepackage{color}
\usepackage{tabularx}
\usepackage{url}

\begin{document}

\title{Higher-order topological superconductors characterized by Fermi level crossings}

\author{Hong Wang}
\affiliation{School of Physics, MOE Key Laboratory for Non-equilibrium Synthesis and Modulation of Condensed Matter, Xi’an Jiaotong University, Xi’an, Shaanxi 710049, China}
\author{Xiaoyu Zhu}
\email{xiaoyu.zhu@xjtu.edu.cn}
\affiliation{School of Physics, MOE Key Laboratory for Non-equilibrium Synthesis and Modulation of Condensed Matter, Xi’an Jiaotong University, Xi’an, Shaanxi 710049, China}

\date{\today}

\begin{abstract}
    We demonstrate that level crossings at the Fermi energy serve as robust indicators for higher-order topology in two-dimensional superconductors of symmetry class D. These crossings occur when the boundary condition in one direction is continuously varied from periodic to open, revealing the topological distinction between opposite edges. The associated Majorana numbers acquire nontrivial values whenever the system supports two Majorana zero modes distributed at its corners. Owing to their immunity to perturbations that break crystalline symmetries, Fermi level crossings are able to characterize a wide range of higher-order topological superconductors. By directly identifying the level-crossing points from the bulk Hamiltonian, we establish the correspondence between gapped bulk and Majorana corner states in higher-order phases. In the end, we illustrate this correspondence using two toy models. Our findings suggest that Fermi level crossings offer a possible avenue for characterizing higher-order topological superconductors in a unifying framework.
\end{abstract}

\maketitle

\section{Introduction}
Topological states of matter are usually endowed with a bulk-boundary correspondence, which facilitates the identifications of topologically protected gapless boundary modes without going into the details of the energy spectrum at open boundaries \cite{hasan2010,qi2011,chiu2016}. Recent advancements in higher-order topological systems have extended this correspondence to include gapped boundaries \cite{benalcazar2017,benalcazar2017a,langbehn2017,song2017,ezawa2018,schindler2018,schindler2018a,khalaf2021,wang2018,zhang2019}, with gapless corner (hinge) modes appearing at the intersections of adjacent edges (surfaces). Tremendous efforts have been devoted to classifying and characterizing these topological states, mostly in crystalline-symmetry protected systems \cite{khalaf2018,geier2018,trifunovic2019,skurativska2020,ono2020,takahashi2020,hsu2020,tang2022,yan2019,zhang2022,bouhon2019,hwang2019,kruthoff2017,tang2019,zhang2019a,vergniory2019,zhang2022a,jung2021,roberts2020,kooi2021,huang2021}. However, it is well known that gapless corner or hinge states persist when crystalline symmetries are broken. This is especially evident in higher-order topological superconductors \cite{zhu2019,yan2018,liu2018,tiwari2020,volpez2019,wu2020}, where Majorana zero modes \cite{read2000,wilczek2009,alicea2012,stanescu2013,elliott2015,Aguado2017} remain stable as mid gap states unless the bulk or boundary gap closes. Hence it would be desirable to characterize higher-order states regardless of whether crystalline symmetries are present.

Higher-order topology can be understood from a boundary perspective, as different parts of the whole boundary, such as the four edges of a square lattice, may exhibit a distinct topology in higher-order phases. For intrinsic higher-order states, the relevant crystalline symmetry requires symmetry related edges or surfaces to be topologically inequivalent \cite{khalaf2018,geier2018,trifunovic2019}. A topology change is only possible through bulk-gap closing. Consequently, bulk invariants, such as symmetry indicators related to the crystalline symmetry, can be defined \cite{skurativska2020,ono2020,takahashi2020,hsu2020}. This stands in contrast with boundary-obstructed topological states, which fall within an extrinsic higher-order classification \cite{khalaf2021,wu2020}. Without the protections of crystalline symmetries, the boundary topology in these states could change while the bulk gap remains open. One may characterize the topology by Wilson loop eigenvalues of Wannier bands that are obtained from Wilson loops of energy bands, the so-called nested Wilson loop approach \cite{benalcazar2017,benalcazar2017a}. However, the quantization of such topological invariants still requires the presence of crystalline symmetries, such as mirror symmetry \cite{khalaf2021,tiwari2020}. Establishing bulk-boundary correspondence under broken crystalline symmetries remains an open question. Considering that boundary topology is ultimately determined by the bulk properties for both intrinsic and boundary-obstructed phases, it should be possible to associate a topological invariant with it based on bulk information, which applies in both phases.

\begin{figure}[t]

    \includegraphics[width=0.48\textwidth]{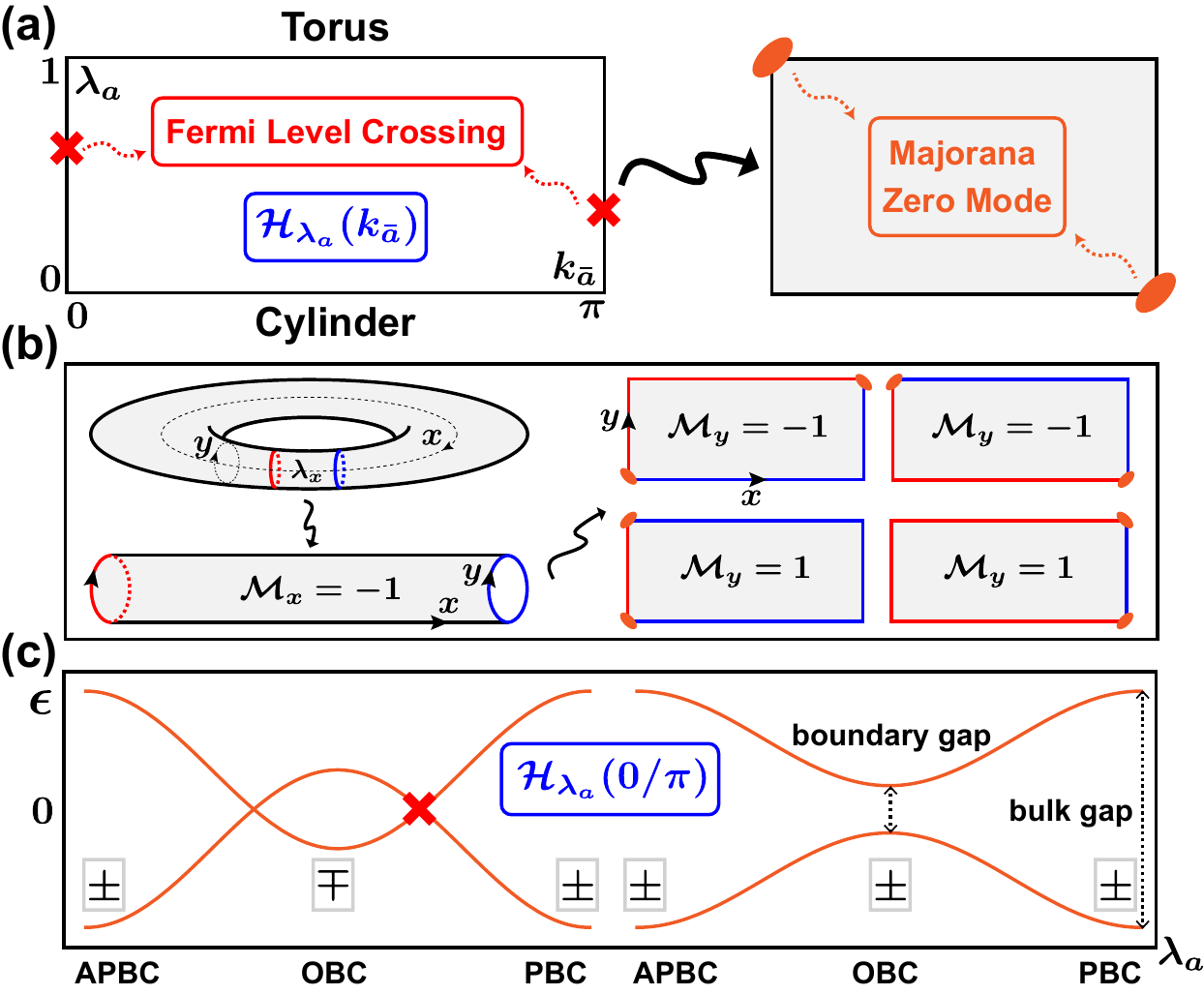}
    \caption{(a) Higher-order topology characterized by Fermi level crossings in the D symmetry class. Crossings occur at high-symmetry momenta $K=0,\pi$, when the system on a torus ($\lambda_a=1$) is continuously deformed into a cylinder ($\lambda_a=0$). (b) Topology of edges determined by Majorana numbers. When $\mathcal M_{x(y)}=-1$, two opposite edges along the $y(x)$ direction exhibit a distinct topology, indicated by different colors. (c) Schematic plots of the BdG spectrum at $K=0 (\pi)$ with or without level crossings. The fermion parity of the ground state (``$+$'' for even, ``$-$'' for odd) switches at each crossing. The crossing may disappear when the bulk or boundary gap closes.}\label{fig1}

\end{figure}

In this paper, we focus on two-dimensional (2D) superconductors of symmetry class D \cite{chiu2016} and higher-order phases featuring two Majorana corner states. The higher-order topology can be characterized by a pair of Majorana numbers, which are intimately related to Fermi level crossings that emerge during the continuous variation of the boundary condition along one direction, as illustrated in Fig. \ref{fig1}(a). We further introduce a generic method for locating these crossings from the bulk Hamiltonian. As a result, bulk-boundary correspondence is established in both higher-order phases discussed earlier, due to the robustness of the Fermi level crossings against crystalline-symmetry-breaking perturbations.

\section{General theory}
To demonstrate how Fermi level crossings determine the higher-order topology of D-class superconductors, we start from a 2D periodic lattice and modulate its boundary condition in one direction. The resulting Bogoliubov-de Gennes (BdG) Hamiltonian can be expressed as
\begin{equation}
    \tilde{\mathcal H}_{\lambda_a} = \sum_{k_{\bar a}}\mathcal H_{\lambda_a}(k_{\bar a}) = \tilde{\mathcal H}_1 - (1-\lambda_a)\sum_{k_{\bar a}}\mathcal B_a(k_{\bar a}),\label{eq:ham2D}
\end{equation}
where $\bar a = y(x)$ when $a=x(y)$, and the real parameter $\lambda_a$ controls the boundary condition in the $a$ direction, with $\lambda_a=1,-1,0$ corresponding to the periodic (PBC), anti periodic (APBC), and open boundary condition (OBC), respectively. In Eq. (\ref{eq:ham2D}), $\mathcal H_{\lambda_a}(k_{\bar a})$ represents the 1D boundary-modulated Hamiltonian at wave vector $k_{\bar a}$, and $\mathcal B_a(k_{\bar a})$ involves all terms that cross its boundary. The lattice terminations we consider are compatible with unit cells, thus allowing the specific form of $\mathcal B$ to be directly read off from the bulk Hamiltonian $\tilde{\mathcal H}_1$. The process of varying $\lambda_a$ from $1$ to $0$ is akin to gradually cutting a torus along the $\bar a$ direction until it eventually becomes a cylinder, as illustrated in Fig. \ref{fig1}(b) for the case of $a=x$.

Here, we consider a gapped bulk with trivial first-order topology, which means the cylindrical system described by $\tilde{\mathcal H}_{\lambda_a=0}$ is fully gapped. Treating it as a quasi-1D system along the $\bar a$-direction, we may characterize the higher-order topology with the Majorana number \cite{kitaev2001,kheirkhah2021,poduval2023}
\begin{equation}
    \mathcal M_a = \text{sgn} \prod_{K} \text{Pf} [-iH_{\lambda_a=0}(K)],\label{eq:MajNumCyl}
\end{equation}
where ``Pf" is shorthand for Pfaffian, $K = 0, \pi$ represents the high-symmetry momentum, and $H$ refers to the matrix representation of $\mathcal H$ in the Majorana basis. In 1D, the Majorana number being $-1$ implies the presence of a single Majorana zero mode at each end. If either $\mathcal M_x$ or $\mathcal M_y$, or both of them, take the value of $-1$, we will instead have two Majorana zero modes at the corners of a 2D sheet. To elaborate this let us consider the cylindrical system in the lower left-hand panel of Fig. \ref{fig1}(b) with $\mathcal M_x=-1$. If we cut it along the axis, the resulting two edges along the $x$ direction will each harbor one Majorana mode. Due to the trivial first-order topology, these localized modes cannot propagate along the edges and must be confined to their respective ends, i.e., the corners. If, in addition $\mathcal M_y = -1$, the two modes would also appear at the two edges in the $y$ direction. As a result, they can only reside at opposite corners, as depicted in the upper right-hand panel of Fig. \ref{fig1}(b). If $\mathcal M_y = 1$, however, they would appear at adjacent corners along the $y$ direction, as shown in the lower right-hand panel of Fig. \ref{fig1}(b).

The Majorana number defined in Eq. (\ref{eq:MajNumCyl}) is closely related to level crossings at the Fermi energy $\epsilon=0$ that appear while $\lambda_a$ varies in the range $[0,1]$. Notably, Eq. (\ref{eq:MajNumCyl}) only involves the 1D Hamiltonian at high-symmetry momenta $K$. Therefore, we only need to consider Fermi level crossings in these subsystems, as shown in Fig. \ref{fig1}(a). At each crossing, the fermion parity of the ground state switches, indicated by the sign change of $\text{Pf}[-iH_{\lambda_a}(K)]$. We can then characterize the fermion-parity difference between PBC and OBC by the number of crossings in between, denoted by $\eta_{a,K}$, as Fig. \ref{fig1}(c) demonstrates. This is formally expressed as
\begin{equation}
    (-1)^{\eta_{a,K}} = \frac{\text{sgn Pf} [-iH_{\lambda_a=0}(K)] }{\text{sgn Pf} [-iH_{\lambda_a=1}(K)] }.\label{eq:crossing}
\end{equation}
We may also define a Majorana number for the toroidal system $\tilde{\mathcal H}_{\lambda_a=1}$ ($\tilde{\mathcal H}_{1}$) similar to Eq.(\ref{eq:MajNumCyl}), which due to trivial first-order topology must be positive, i.e.,
\begin{equation}
    \text{sgn} \prod_{K} \text{Pf} [-iH_{\lambda_a=1}(K)] = 1.\label{eq:MajNumCyl1}
\end{equation}
Combining Eqs. (\ref{eq:MajNumCyl})-(\ref{eq:MajNumCyl1}), we arrive at
\begin{equation}
    \mathcal M_a = \prod_{K}(-1)^{\eta_{a,K}}=(-1)^{\eta_a}, \label{eq:MajNum2D}
\end{equation}
where $\eta_a$ denotes the total number of crossings at $K=0,\pi$. An odd value of $\eta_x$ or $\eta_y$ implies the system resides in a higher-order phase. Fermi level crossings are protected by fermion-parity conservation and particle-hole symmetry, making them immune to crystalline-symmetry-breaking perturbations \cite{beenakker2013}.

Intuitively, we may understand the relation between Fermi level crossings and higher-order topology from the viewpoint of boundary topology. As shown in Fig. \ref{fig1}(b), an odd value of $\eta_a$ ($\mathcal M_a=-1$) reveals that opposite edges along $\bar a$ are topologically inequivalent (shown in different colors). This explains the possible locations of Majorana zero modes, which appear at the intersections of topologically distinct edges. In some simple models, as we demonstrate later, the edge topology can be characterized by the sign of the mass gap in the edge Hamiltonian, allowing us to validate this argument.

To establish the bulk-boundary correspondence, we will demonstrate how the Fermi level crossings of the 1D subsystems are identified from the bulk Hamiltonian. For brevity, we use $\mathcal H_\lambda$ to replace $\mathcal H_{\lambda_a}(K)$, where
\begin{equation}
    \mathcal H_\lambda  = \mathcal H_1 - (1-\lambda)\mathcal B \label{eq:ham1D}
\end{equation}
represents a generic 1D Hamiltonian of D class. Following the prescription given by Ref.\cite{rhim2018}, we first define a retarded Green's function
\begin{equation}
    \mathcal G_\lambda(\epsilon) = (\epsilon - \mathcal H_\lambda +i\tilde\delta)^{-1}= A^{-1}_\lambda(\epsilon)\mathcal G_1(\epsilon)\mathcal,
\end{equation}
where $\tilde\delta$ is a positive infinitesimal, $\mathcal G_1(\epsilon)$ is the Green's function corresponding to the bulk Hamiltonian $\mathcal H_1$, and $\mathcal A_\lambda(\epsilon) = 1+(1-\lambda)\mathcal G_1(\epsilon)\mathcal B$. Since we focus on the parameter regime in which the bulk is fully gapped, in-gap states of $\mathcal H_{\lambda}$ are solely determined by the poles of $\mathcal A^{-1}_\lambda(\epsilon)$. Consequently, level-crossing points are identified as roots of 
\begin{equation}
    \det [A_\lambda(\epsilon=0)] = 0, \label{eq:detA}
\end{equation}
where $A_\lambda$ is the matrix representation of $\mathcal A_\lambda$. As $\mathcal B$ only includes intra-cell terms crossing the boundary, we then have $[A_\lambda]_{IJ}=\delta_{IJ}$ if $J$ does not appear in these terms. This enables us to calculate $\det (A)$ using a much smaller matrix $D$, which is obtained by projecting $\mathcal A$ into the eigenspace of $\mathcal B$ and satisfies $\det (D)=\det (A)$. The entries of $D$ are given by
\begin{equation}
    [D_\lambda(\epsilon)]_{ij} = \delta_{ij}+(1-\lambda)\sum_{n,k}\frac{\langle i|n,k\rangle\langle n,k|\mathcal B|j\rangle}{\epsilon-\epsilon_{n,k}},\label{eq:matD}
\end{equation}
where $\epsilon_{n,k}$ denotes the energy spectrum of the bulk Hamiltonian $\mathcal H_1$ in the Brillouin zone, with $|n,k\rangle$ being the corresponding eigenstate, and $|i\rangle$,$|j\rangle$ represent the eigenvectors of $\mathcal B$. The dimension of $D_\lambda$ is equal to the rank of $\mathcal B$, denoted by $N_b$. We then obtain the characteristic equation
\begin{equation}
    \det[D_\lambda(\epsilon=0)] = 0, \label{eq:detD}
\end{equation}
which has $N_b$ roots in total. The number of Fermi level crossings $\eta$ is half the number of real roots in the interval $[0,1]$, from which we can readily obtain Majorana numbers according to Eq. (\ref{eq:MajNum2D}).

Compared to Eq.(\ref{eq:MajNumCyl}), where Majorana numbers are determined by calculating the Pfaffian of finite systems with open boundaries \cite{wimmer2012}, i.e., $\text{Pf} [-iH_{\lambda_a=0}(K)]$, and the accuracy crucially depends on system size, identifying the Fermi level crossings is computationally more accurate and efficient for a translation-invariant system. It does not suffer from finite-size effects, and the computational cost is similar to Wilson loop calculations. Moreover, it provides a potential path to characterizing higher-order topological superconductors in other symmetry classes such as the DIII or BDI classes, where Fermi level crossings might be protected by their topological charges. Additionally, by pinpointing the crossings directly from the bulk Hamiltonian, we establish the correspondence between gapped bulk and gapless corner states in higher-order phases. In the following, we shall illustrate this in specific models.

\begin{figure}[t]

    \includegraphics[width=0.48\textwidth]{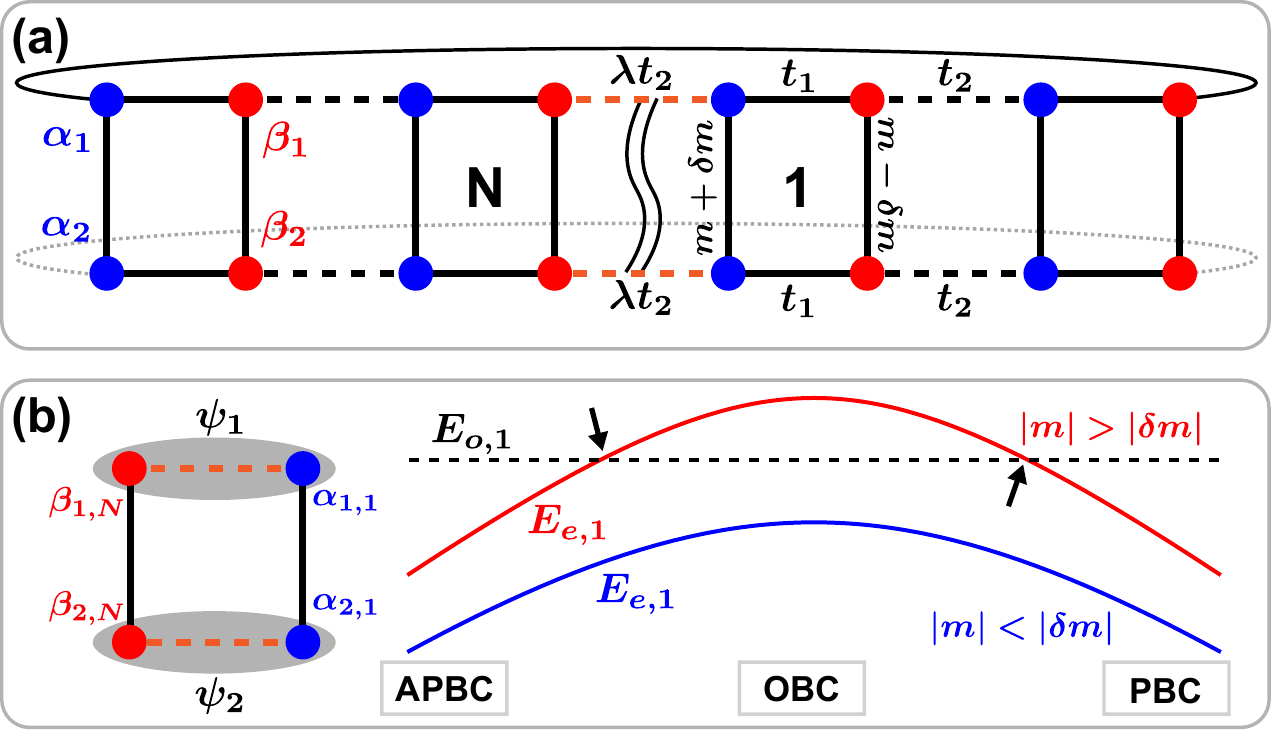}
    \caption{(a) Geometry of the two-leg Kitaev ladder. $\lambda$ controls the boundary condition. (b) The fermion-parity switch in a dimerized lattice ($t_1=0$). For the case of $|m|>|\delta m|$, the ground state switches from the even-parity sector ($E_{e,1}$, solid lines) to the odd-parity sector ($E_{o,1}$, dashed lines) while the boundary condition varies from PBC (APBC) to OBC. Level crossings are indicated by black arrows.} \label{fig2}

\end{figure}

\section{toy models}
First we consider a two-leg Kitaev ladder \cite{wu2012,chitov2018,wakatsuki2014,yan2020} as schematically shown in Fig. \ref{fig2}(a), and demonstrate how level crossings are identified from the bulk Hamiltonian. Each unit cell contains four Majorana fermions denoted by $\alpha_{s,j}$ and $\beta_{s,j}$, with $s=1,2$ being the chain index and $j$ referring to the cell index. The boundary-modulated Hamiltonian with $N$ unit cells has the form $\mathcal H_\lambda = \Gamma^T H_\lambda \Gamma$ in the Majorana basis $\Gamma = \bigoplus_{j=1}^N \Gamma_j$, where $\Gamma_j = \{\alpha_{1,j},\alpha_{2,j},\beta_{1,j},\beta_{2,j}\}^T/\sqrt{2}$ and the Hamiltonian matrix is given by
\begin{equation}
    H_\lambda = \sum_{r=0,\pm 1} T^r\otimes h_r + \lambda(T^{N-1}\otimes h_1^\dagger + \text{H.c.}).  \label{eq:ham}
\end{equation}
Here, $T$ denotes the translation operator that moves each cell by one site to the left, with $T|j\rangle=|j-1\rangle$ and $T|j=1\rangle = 0$ \cite{alase2016}. Hamiltonian (\ref{eq:ham}) includes the intra cell term $h_0 = -t_1\tau_y-m \sigma_y-\delta m \tau_z\sigma_y$, and inter cell term $h_1 = h_{-1}^\dagger = t_2 (\tau_y+i\tau_x)/2$, with $\tau$ and $\sigma$ being Pauli matrices that act in the chain and rung space separately. $t_1$ and $t_2$ represent couplings of Majorana fermions along the chain, while $m$ and $\delta m$ are those along the rung. For brevity, we assume $t_1$ and $t_2$ to be non-negative.

In this model, $m$ and $\delta m$ determine whether level crossings occur when $\lambda$ varies in the range $[0,1]$. This is readily seen in a perfectly dimerized lattice ($t_1=0$), in which case only the boundary block shown in Fig. \ref{fig2}(b) depends on $\lambda$, and its Hamiltonian has the form
\begin{align}
    \mathcal H_b = &-2\lambda t_2(\psi_1^\dagger\psi_1+\psi_2^\dagger \psi_2-1) \nonumber\\
                   &+2i (m\psi_1^\dagger\psi_2+\delta m\psi_1^\dagger \psi_2^\dagger - \text{H.c.}),
\end{align}
where $\psi_s = (\alpha_{s,1}+i \beta_{s,N})/2$ are fermionic operators. The conservation of fermion number parity enables us to study the lowest energy levels in the even- and odd-parity sectors separately, with $E_{e,1} = -2\sqrt{\lambda^2 t_2^2+\delta m^2}$ and $E_{o,1} = -2|m|$. While the boundary condition goes from PBC to OBC, the two levels would cross if $0<m^2-\delta m^2<t_2^2$, signaling a switch in the ground-state fermion parity, as demonstrated in Fig. \ref{fig2}(b). This parity switch could be observed from the zero-bias peak in an experimental setup that consists of two quantum dots coupled by a nanowire-superconductor heterojunction \cite{dvir2023,bordin2023}. The parameters $m$ and $\delta m$ are related to the electrochemical potential of quantum dots, and $t_2$ or $\lambda$ is controlled by tuning the cross Andreev reflection and elastic cotunnelling.

\begin{figure}[t]

    \includegraphics[width=0.48\textwidth]{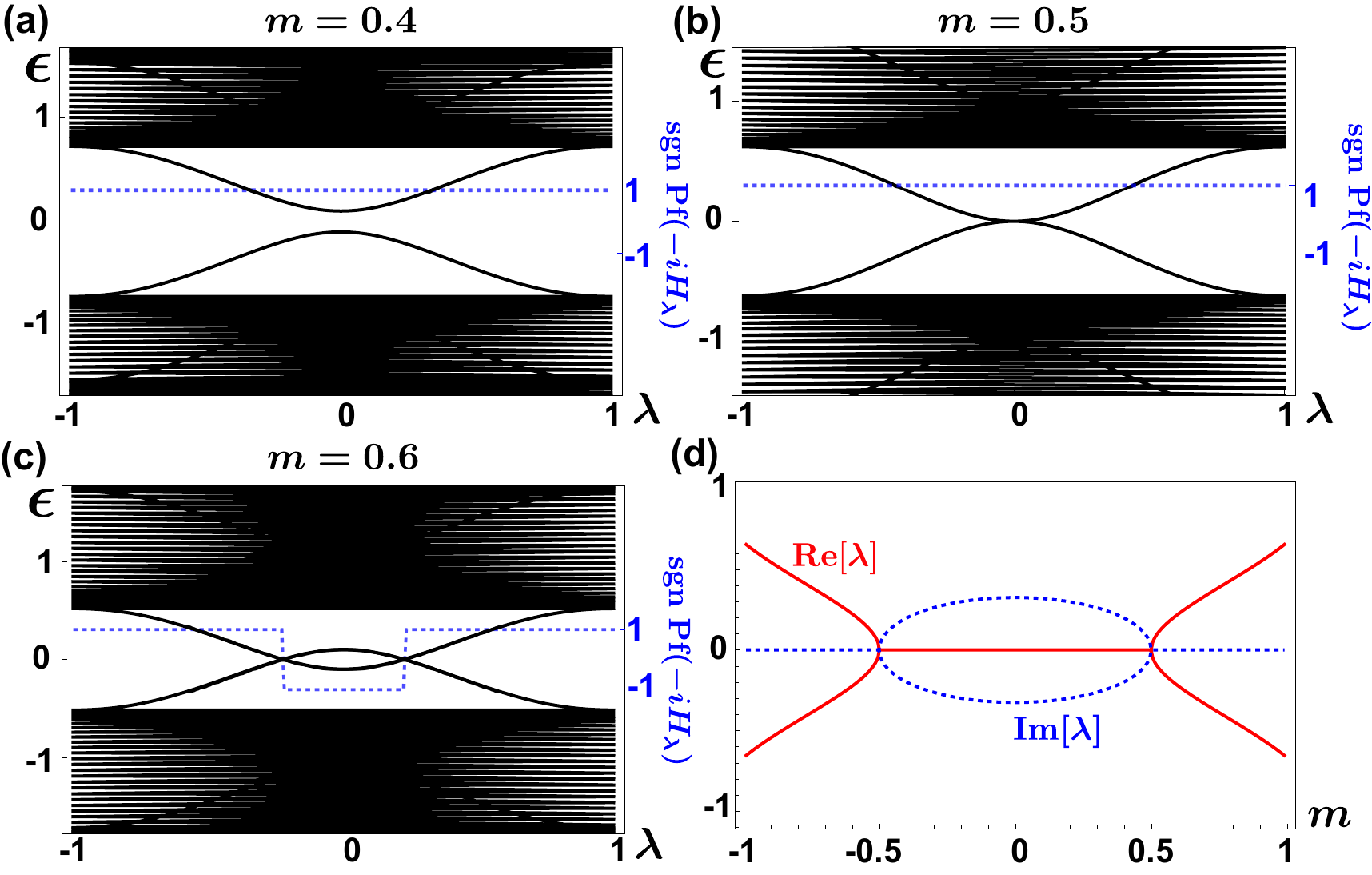}
    \caption{(a)-(c) Fermi level crossings in a BdG spectrum. The fermion parity switches at each crossing. The crossing appears when $|m|>|\delta m|$. Blue dashed lines indicate signs of the Pfaffian of a 1D Hamiltonian. (d) The variation of crossing points $\lambda$ with $m$. $\delta m=0.5$, $t_1=1$, and $t_2=2$.} \label{fig3}

\end{figure}

For generic $t_1$, we have $\mathcal H_{1}=\sum_k\Gamma_{-k}^T H_k \Gamma_k$ in $k$ space, with the basis $\Gamma_k = \{\alpha_{1,k},\alpha_{2,k},\beta_{1,k},\beta_{2,k}\}^T/\sqrt{2}$, and the Bloch Hamiltonian
\begin{equation}
    H_k = (-t_1+t_2\cos k)\tau_y - t_2 \sin k \tau_x - m\sigma_y - \delta m \tau_z\sigma_y. \label{eq:hamK}
\end{equation}
The energy spectrum is given by
\begin{equation}
    \epsilon_{n,k} = \pm \left(\sqrt{t_1^2+t_2^2-2t_1t_2\cos k+\delta m^2}\pm m\right), \label{eq:spectrum}
\end{equation}
with $n$ being the band index. Substituting $\epsilon_{n,k}$ and $\mathcal B = \Gamma_N^T h_1 \Gamma_1 + \text{H.c.}$ into Eqs. (\ref{eq:matD}) and (\ref{eq:detD}), we obtain
\begin{equation}
    \lambda^2 = 1-\frac{2\Lambda}{\Lambda-a+2t_2^2}, \label{eq:root}
\end{equation}
with $\Lambda=\sqrt{a^2-b^2}$, $a=t_1^2+t^2_2+\delta m^2-m^2$, and $b=2t_1t_2$. From Eq. (\ref{eq:root}), we find that the number of crossings $\eta=1$ when $0<m^2-\delta m^2<(t_2-t_1)^2$ and $t_1<t_2$, as shown in Fig. \ref{fig3}. This indicates that the boundary phase transition occurs at $|m|=|\delta m|$ as in the dimerized case, which is verified by the exact boundary spectrum (see Supplemental Material \cite{supmat} and Ref. \cite{pershoguba2012} therein). In the special case where $\delta m=0$, Hamiltonian (\ref{eq:hamK}) is invariant under inversion, with the corresponding operator being $\tau_y$, up to a gauge factor. The inversion symmetry facilitates the direct determination of the Fermi level crossings from the differences of the ground-state inversion eigenvalues between PBC and APBC \cite{supmat}. With the knowledge of $\eta$ in a 1D system, we can proceed to determine the higher-order topology in a 2D system, according to Eq. (\ref{eq:MajNum2D}).

\begin{figure}[t]

    \includegraphics[width=0.48\textwidth]{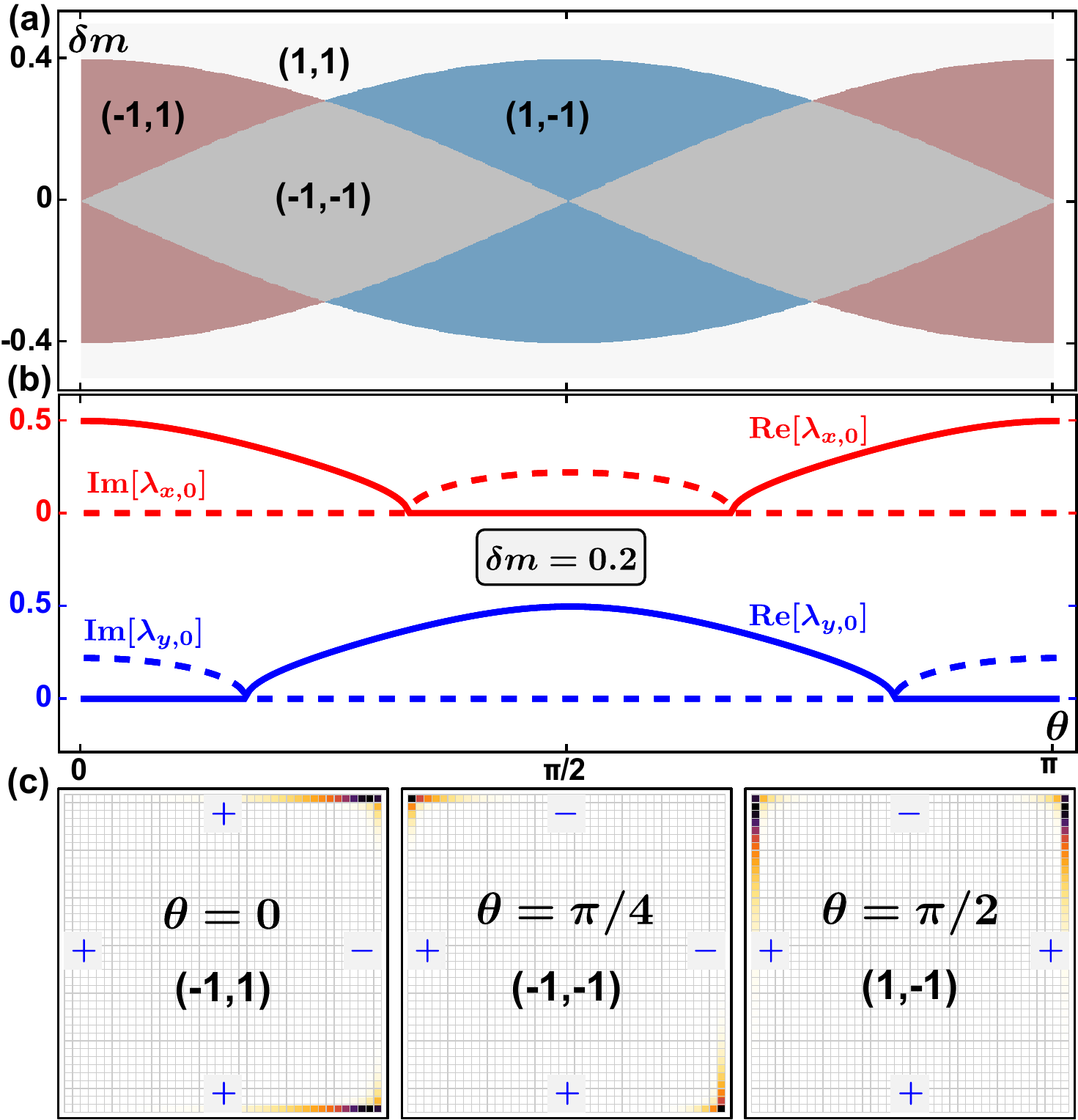}
    \caption{(a) Phase diagram of the 2D model in $(\theta,\delta m)$ space. Four different phases are characterized by Majorana numbers $(\mathcal M_x,\mathcal M_y)$. (b) Evolutions of crossing points with $\theta$. No crossings appear at $K=\pi$ in this case and hence only those at $K=0$ are displayed. (c) Distributions of Majorana zero modes in three nontrivial phases that are separated by boundary phase transitions. Majorana zero modes appear either at two adjacent corners ($\theta=0,\pi/2$), or at opposite corners ($\theta=\pi/4$). The symbols ``$+$'' and ``$-$'' indicate the signs of the edge gaps. $t_1=0.5$, $t_2=1$, and $m=0.4$.}\label{fig4}

\end{figure}

The 2D Hamiltonian we consider takes the form
\begin{align}
    &H^{\text{2D}}_{\bm k} = [t_2(\cos k_x+\cos k_y)-t_1-t_2]\tau_y-\delta m\tau_z\sigma_y  \label{eq:hamK2D}\\
    &-t_2(\sin k_x \tau_x+\sin k_y \tau_z\sigma_z)- m(\cos\theta \sigma_y+\sin\theta \tau_y\sigma_x),\nonumber 
\end{align}
when written in the Majorana basis as in Eq.(\ref{eq:hamK}), and reduces to the 1D Hamiltonian at $k_y=0$, $\theta=0$. This model is equivalent to the $p\pm ip$ superconductor under an in-plane Zeeman field \cite{phong2017,zhu2018}. According to Eq. (\ref{eq:MajNum2D}), Majorana numbers $(\mathcal M_x,\mathcal M_y)$ are determined by Fermi level crossings of four 1D Hamiltonians, $\mathcal H_{\lambda_a}(K)$. In Fig. \ref{fig4}(a), we draw the $(\theta,\delta m)$ phase diagram. Here the crossings only occur at $K=0$ as Fig. \ref{fig4}(b) shows, although it is possible they emerge at $K=\pi$ for $t_1$ and $t_2$ taking other values. Two Majorana corner states emerge when at least one Majorana number takes $-1$, as illustrated in Fig. \ref{fig4}(c).

To corroborate previous arguments concerning the relation between level crossings and boundary topology, we obtain the mass gap for an arbitrary edge \cite{supmat}, given by
\begin{equation}
    \Delta(\phi)=\delta m - m\cos(\phi-\theta), \label{eq:gap}
\end{equation}
where $\phi$ indicates the normal direction of the edge ($\phi=0,\pi/2$ for right and top edges respectively). The topology of the edges in D-class systems can be characterized by the sign of the mass gap. As seen from the three representative cases in Fig. \ref{fig4}(c), gaps of opposite edges along $y(x)$ indeed take different signs when $\eta_{x(y)}$ is an odd number, or equivalently, $\mathcal M_{x(y)}=-1$. This can be guaranteed when inversion symmetry is enforced, by noting that $\Delta(\phi)=-\Delta(\phi+\pi)$ in the absence of $\delta m$. In this intrinsic higher-order phase, we always have $\mathcal M_x = \mathcal M_y = -1$. Turning on $\delta m$ breaks inversion symmetry and drives the system into a boundary-obstructed phase, in which process the gap signs do not change immediately, so is the number of Fermi level crossings. We can therefore use Fermi level crossings to characterize the higher-order topology in both phases.

The robustness of Fermi level crossings is also reflected in their persistence under weak disorder or boundary impurities \cite{supmat}. While Eqs. (\ref{eq:MajNum2D}) and (\ref{eq:detD}) may not be directly applicable due to potential broken translation symmetry, the number of Fermi level crossings remains unchanged. This reinforces their role as a reliable tool to characterize higher-order topological superconductors.

\section{conclusion}
In conclusion, Fermi level crossings can serve as useful indicators for higher-order topology in the D symmetry class when the nontrivial phase accommodates two Majorana corner states. The applicability of this approach extends beyond the toy models introduced above, as demonstrated in the Supplemental Material \cite{supmat} for a Rashba bilayer system. The level crossings we consider emerge while the boundary condition continuously varies from PBC to OBC, during which two opposite edges gradually decouple. An odd number of crossings signals a topological distinction between the two edges. From this point of view, one may consider Fermi level crossings emerging under variations of other twisted boundary conditions \cite{song2020} when dealing with higher-order phases with four or more Majorana corner states, where one needs to associate the crossings with topological distinctions between neighboring edges.

\section*{Acknowledgements}
This work was supported by National Science Foundation of China (NSFC) under Grant No. 11704305, and the Innovation Program for Quantum Science and Technology (2021ZD0302400).

\bibliography{tlkm.bib}

\onecolumngrid

\clearpage

\setcounter{equation}{0}
\setcounter{figure}{0}
\setcounter{page}{1}

\begin{center}
    {\textbf{\large Supplemental Material for ``Higher-order topological superconductors \\[0.2cm] characterized by Fermi level crossings"}\\[1cm]}
\end{center}

\twocolumngrid
\appendix

In this Supplemental Material, we provide detailed derivations of the boundary spectrum for the two-leg Kitaev ladder, explore the role of inversion symmetry in determining Fermi level crossings, derive the effective Hamiltonian for an arbitrary edge in the 2D model, and investigate the robustness of Fermi level crossings against bulk disorder and boundary impurities. We also demonstrate in a Rashba bilayer superconducting system how Fermi level crossings effectively identify higher-order topological phases.

\section{A. Boundary spectrum}

We consider a semi-infinite system with boundary at $j=1$ in the two-leg Kitaev ladder. In the Hamiltonian $H_{\lambda=0}$, $\sigma_y$ is a good quantum number, and we can work in its eigenspace, where $H_{\lambda=0}$ is block diagonal. Consequently, we can set $\sigma_y$ to be $\pm 1$ in the two blocks, respectively. In this toy model, each block with $\sigma_y=1$ ($\sigma_y=-1$) can be viewed as a particle (hole) version of Su-Schrieffer-Heeger (SSH) model, with the parameter $m$ acting as a chemical potential term that shifts the energy spectrum in corresponding block. The two blocks do not couple due to the conservation of $\sigma_y$ in this simple model. It is possible to introduce additional terms that couple the two blocks and make the model more complicated. However, the main results do not change as we only require particle-hole symmetry. The simplicity of this toy model allows us to obtain analytical results in a straightforward manner, as we demonstrate in the following.

We will now derive the condition for the appearance of gapped boundary modes in the two-leg Kitaev ladder, as well as the boundary spectrum. Let's first consider the block with $\sigma_y=1$, and the case with $\sigma_y=-1$ can be obtained by sending $m\rightarrow -m$ and $\delta m\rightarrow -\delta m$. The Hamiltonian matrix with $\sigma_y=+1$ is given by
\begin{equation}
    H = \sum_{r=0,\pm 1} T^r\otimes \tilde h_{r},  \label{eqS:ham}
\end{equation}
where $\tilde h_{0} = -t_1\tau_y-m-\delta m \tau_z$, $\tilde h_{1} = \tilde h_{-1}^\dagger = t_2 (\tau_y+i\tau_x)/2$ and $T$ is the translation operator that moves each unit cell by one site to the left, with $T|j\rangle=|j-1\rangle$ and $T|j=1\rangle = 0$. The wavefunction $\Psi = \bigoplus_{j=1} \phi_j = \bigoplus_{j=1} \{u_j,v_j\}^T$ satisfies the Schr\"{o}dinger equation $H\Psi=E\Psi$, which has the form
\begin{equation}
    (\tilde h_{0}-E)\phi_{1} + \tilde h_{1}\phi_{2} = 0 \label{eqS:s1}
\end{equation}
and
\begin{equation}
    \tilde h_{1}^\dagger\phi_{j-1}+(\tilde h_{0}-E)\phi_{j}+\tilde h_{1}\phi_{j+1}=0 \label{eqS:s2}
\end{equation}
for $j>1$. Multiplying Eq.(\ref{eqS:s2}) by $q^{j-1}$ and summing up all the equations, we obtain
\begin{equation}
    [q \tilde h_1^\dagger + (\tilde h_0-E)+q^{-1}\tilde h_1]G(q) = (\tilde h_0-E+\tilde h_1 q^{-1})\phi_1 + \tilde h_1\phi_2 \label{eqS:s3}
\end{equation}
where 
\begin{equation}
    G(q) = \sum_{j=1}^{+\infty}q^{j-1}\phi_j,
\end{equation}
and $q$ is a complex number. Utilizing Eq.(\ref{eqS:s1}) and (\ref{eqS:s3}), we can express $G(q)$ as 
\begin{equation}
    G(q) = [q^2 \tilde h_1^\dagger + q (\tilde h_0-E)+\tilde h_1]^{-1}\tilde h_1\phi_1. \label{eqS:G1}
\end{equation}
For $\Psi$ to be a localized state at $j=1$, all the poles of $G(q)$ must satisfy $|q_p|>1$ \cite{pershoguba2012}. Substituting the specific forms of $\tilde h_0$ and $\tilde h_1$ into Eq.(\ref{eqS:G1}), we have
\begin{equation}
    G(q) = \frac{u_1t_2}{D}\biggl \{t_1-qt_2 ,-i(m+\delta m + E)\biggl \}^T,\label{eqS:G2}
\end{equation}
with the denominator
\begin{equation}
    D = (t_1-qt_2)(t_2-qt_1)+q[(E+m)^2-\delta m^2].
\end{equation}
The poles are decided from the two roots of $D=0$, which have the relation $q_1q_2=1$, and therefore cannot both have absolute values greater than one. So only one of the two roots can be the pole and the other has to be eliminated from numerator. By requiring $(E+m)^2-\delta m^2=0$, we could eliminate $t_1-qt_2$ term in the first entry of Eq.(\ref{eqS:G2}). However, only when $E = -m-\delta m$ can this term be eliminated in both entries, which leads to 
\begin{equation}
    G(q) =  \frac{u_1 t_2}{t_2-qt_1}\{1,0\}^T.
\end{equation}
The pole $q_p = t_2/t_1$, satisfies $|q_p|>1$ when $t_1<t_2$. A series expansion of $G(q)$ takes the form
\begin{equation}
    G(q) = \sum_{j=1}^{+\infty} q^{j-1}\left(\frac{t_1}{t_2}\right)^{j-1}\{u_1,0\}^T.
\end{equation}
Comparing this equation with the definition of $G(q)$, we immediately find that
\begin{equation}
    \phi_j = \left(\frac{t_1}{t_2}\right)^{j-1}\{u_1,0\}^T,
\end{equation}
which is clearly localized at $j=1$ if $t_1<t_2$. For the block with $\sigma_y=-1$, the energy of bound state is given by $E=m+\delta m$. Hence the boundary mode at $j=1$ appears when $t_1<t_2$. The boundary states for $H_{\lambda=0}$ at $j=1$ can be written in original Majorana basis, with
\begin{align}
    &E_{b1,+} = m+\delta m, \nonumber\\
    &\Psi_{b1,+} = \frac{u_1}{\sqrt{2}} \bigoplus_{j=1} \left(\frac{t_1}{t_2}\right)^{j-1}\{1,-i,0,0\}^T  \\
    &E_{b1,-} = -(m+\delta m), \nonumber\\
    &\Psi_{b1,-} = \frac{u_1}{\sqrt{2}} \bigoplus_{j=1} \left(\frac{t_1}{t_2}\right)^{j-1}\{1,i,0,0\}^T
\end{align}
where $u_1$ is decided from normalization condition.

Boundary modes at the other end can be obtained in a similar way. We consider a semi-infinite system with boundary at $j=N$. Schr\"{o}dinger equation for $\sigma_y=1$ block is given by
\begin{equation}
    \tilde h_{1}^\dagger\phi_{N-1} + \tilde h_{0}\phi_{N}= E\phi_{N} \label{eqS:sN1}
\end{equation}
and
\begin{equation}
    \tilde h_{1}^\dagger\phi_{j-1}+(\tilde h_{0}-E)\phi_{j}+\tilde h_{1}\phi_{j+1}=0 \label{eqS:sN2}
\end{equation}
for $j\in(-\infty,N-1]$. Multiplying Eq.(\ref{eqS:sN2}) by $q^{N-j}$ and summing up all the equations, we have 
\begin{equation}
    G(q) = \sum_{j=N}^{-\infty}q^{N-j}\phi_j = [\tilde h_1^\dagger+q(\tilde h_0-E)+q^2\tilde h_1]^{-1}\tilde h_1^\dagger \phi_N
\end{equation}
Following the same analysis in deriving boundary modes at $j=1$, we obtain the boundary states at $j=N$, with
\begin{align}
    &E_{bN,+} = m-\delta m, \nonumber\\
    &\Psi_{bN,+} = \frac{v_N}{\sqrt{2}} \bigoplus_{j=N}^{-\infty} \left(\frac{t_1}{t_2}\right)^{N-j}\{0,0,1,-i\}^T \\
    &E_{bN,-} = -(m-\delta m), \nonumber \\
    &\Psi_{bN,-} = \frac{v_N}{\sqrt{2}} \bigoplus_{j=N}^{-\infty} \left(\frac{t_1}{t_2}\right)^{N-j}\{0,0,1,i\}^T.
\end{align}

So we have established that gapped boundary modes appear when $t_1<t_2$, and the boundary spectrum is independent of $t_1$, being $\pm(m+\delta m)$ at the end $j=1$, and $\pm(m-\delta m)$ at $j=N$. Note that, these boundary modes may appear in the bulk continuum if bulk gap vanishes or is smaller than the boundary gap. For a gapped bulk, boundary phase transition occurs when $|m|=|\delta m|$. In this context, we can treat each boundary as a zero-dimensional gapped system that belongs to D class, which is again characterized by a $\mathbb Z_2$ invariant \cite{chiu2016}. One can in principle assign a different invariant at two sides of the phase transition point based on the number of Fermi level crossings. However, this is not related to gapless modes.

\section{B. Inversion symmetry}

\begin{figure}[t]
    \includegraphics[width=0.48\textwidth]{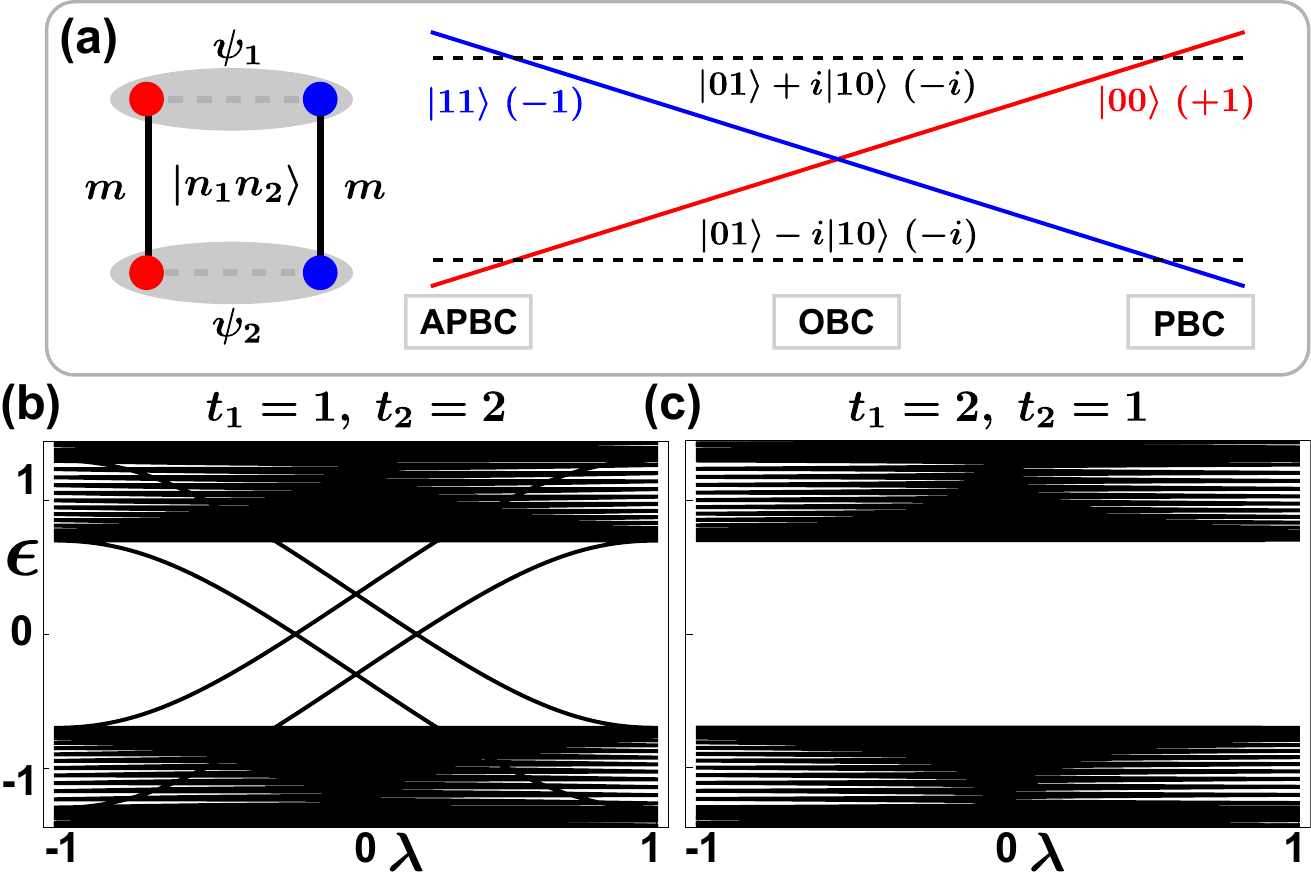}
    \caption{(a) Fermion-parity switch with inversion symmetry enforced ($\delta m=0$). Although the ground state falls in even-parity sector for both PBC and APBC, they have distinct inversion eigenvalues (shown in the parenthesis). (b) For $t_1<t_2$ the spectrum exhibits spectral flows with some states moving from occupied (negative energies) to unoccupied (positive energies) bands. (c) Due to inversion symmetry, level crossings can only disappear when bulk gap closes and reopen.} \label{figs1}
\end{figure}

In this section, we demonstrate that with inversion symmetry enforced, level crossings can be inferred from ground-state difference of inversion eigenvalues for system under PBC and APBC. 

In the absence of $\delta m$, the 1D Kitaev ladder is invariant under inversion, which transforms Majorana fermions as
\begin{equation}
    \mathcal I\alpha_{s,j}\mathcal I^{-1} = -\beta_{s,N+1-j}, \ \mathcal I\beta_{s,j}\mathcal I^{-1} = \alpha_{s,N+1-j}. \label{eq:inversion}
\end{equation}
Accordingly, fermionic operators $\psi_s$ follow transformation $\mathcal I\psi_s \mathcal I^{-1} = i \psi_s$. Although fermion parity is still the same for PBC and APBC, we could discriminate the two ground states by inversion eigenvalues. Specifically, for $t_1=0$, provided $m$ is finite, the ground state would evolve from $|11\rangle = \psi_1^\dagger\psi_2^\dagger|00\rangle$ under PBC to $|00\rangle$ in APBC. The inversion eigenvalues of the two states are $-1$ and $1$ respectively. This distinction leads to level crossings, as shown in Fig. \ref{figs1}(a). 

As Fig. \ref{figs1}(b) shows, the level crossings persist for finite $t_1$, with two states from occupied bands (negative energy) moving straight into unoccupied bands (positive energy) as $\lambda$ varies. This suggests that ground state under APBC should be different from that of PBC in some aspect. As we illustrate in Fig. \ref{figs1}(a) for $t_1=0$, the difference lies in their inversion eigenvalues. To investigate this for general $t_1$, we may write down the mean-field ground state explicitly. Define fermionic operators $c_{s,k} = (\alpha_{s,k} + i\beta_{s,k})/2$, and we could express the 1D Hamiltonian in particle-hole basis $\{c_{1,k},c_{2,k},c_{1,-k}^\dagger,c_{2,-k}^\dagger\}^T$, which takes the form
\begin{equation}
    \tilde{H}_k = (t_1-t_2 \cos k)\tilde{\tau}_z + t_2 \sin k \tilde{\tau}_y - m\tilde{\sigma}_y,\label{eq:BdG}
\end{equation}
with $\tilde{\tau}$ and $\tilde{\sigma}$ being Pauli matrices acting in Nambu space and chain space. Ground state of this Bogoliubov-de Gennes (BdG) Hamiltonian is given by \cite{read2000}
\begin{equation}
    |G\rangle = \mathcal N \exp \left(\sum_{k>0, s}g_k c_{s,k}^\dagger c_{s,-k}^\dagger\right) \prod_{n\in\text{occ},K}(\phi_{n,K}^\dagger)^{\nu_{n,K}} |0\rangle, \label{eq:groundState}
\end{equation}
where $\mathcal N$ is a normalization factor, $g_k = it_2\sin k/[\epsilon_0+(t_1-t_2\cos k)]$ with $\epsilon_0 = \sqrt{t_1^2+t_2^2-2t_1t_2\cos k}$, and $\phi_{n,K}^\dagger$ is the creation operator of Bogoliubov quasiparticle in occupied bands at high symmetry momenta. Occupation number $\nu_{n,K}$ is determined by the eigenstate of $\phi_{n,K}$, denoted by $|\widetilde{n,K}\rangle = \{u_{n1,K},u_{n2,K},v_{n1,K},v_{n2,K}\}^T$, where $u_{ns,K}$ and $v_{ns,K}$ are particle and hole components respectively. We would then have $\phi_{n,K}^\dagger = \sum_s u_{ns,K}c_{s,K}^\dagger+v_{ns,K}c_{s,K}$. At $K$, there is no pairing term. Only the state that is of particle type ($v_{ns,K}=0$) will be occupied in the ground state, with $\nu_{n,K}=1$.

According to Eq.(\ref{eq:inversion}), $\mathcal I c_{s,k} \mathcal I^{-1} = i c_{s,-k}e^{-ik(N+1)}$, which is differing by a sign for PBC and APBC. Inversion symmetry requires BdG Hamiltonian (\ref{eq:BdG}) to obey $U_I^\dagger(-k) \tilde{H}(k) U_I(-k) = \tilde H(-k)$, with $U_I(k) = i e^{ik}\tilde{\tau}_z$. Each state at $K$ would be an eigenstate of $\tilde\tau_z$, of which particle states satisfying $\tilde{\tau}_z|\widetilde{n,K}\rangle = |\widetilde{n,K}\rangle$. Therefore, $\nu_{n,K}$ is identified as the number of occupied states $|\widetilde{n,K}\rangle$ with $\tilde\tau_z = 1$. Under inversion transformations, pairing terms in Eq.(\ref{eq:groundState}) remain the same while $\phi_{n,K}^\dagger$ acquires a factor of $-ie^{iK}$ (under PBC). The difference of ground-state inversion eigenvalues between PBC and APBC can then be given by
\begin{equation}
    \mathcal P = \prod_{n\in\text{occ}}(-i)^{\nu_{n,0}-\nu_{n,\pi}},\label{eq:invGround}
\end{equation}
which is valid regardless of $N$ being even or odd. $\mathcal P$ can take four different values and therefore serves as a $\mathbb Z_4$ invariant. Ground-state fermion-parity difference between PBC and APBC may also be expressed with the occupation numbers, given by
\begin{equation}
    n_F = \prod_{n\in\text{occ},K}(-1)^{\nu_{n,K}}.
\end{equation}
For $\mathcal P\neq 1$, the spectrum exhibits spectral flows while $\lambda$ varies, with some states moving from occupied bands to unoccupied bands, and therefore level crossing is inevitable. When $\mathcal P=\pm i$, there would be an odd number of level crossings while $\lambda$ varies between PBC and APBC due to the fermion-parity difference ($n_F=-1$), thus leaving an unpaired Majorana zero mode at each open boundary. The gapped boundaries with nontrivial topology is characterized by $\mathcal P = -1$.

\section{C. Effective Edge Hamiltonian}

\begin{figure}
    \includegraphics[width=0.3\textwidth]{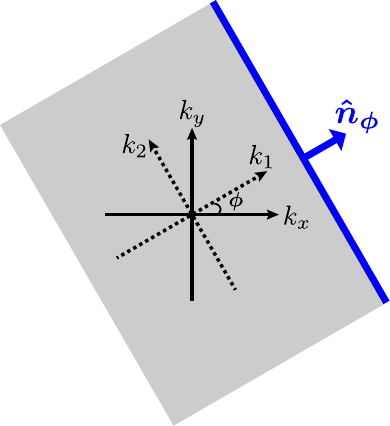}
    \caption{A semi-infinite 2D system with boundary (blue line) along $k_2$ direction. Two sets of coordinate systems are related by an in-plane rotation. $\hat n$ represents the unit vector pointing along normal direction of the edge.}\label{figs2}
\end{figure}

\begin{figure*}[t]
    \includegraphics[width=\textwidth]{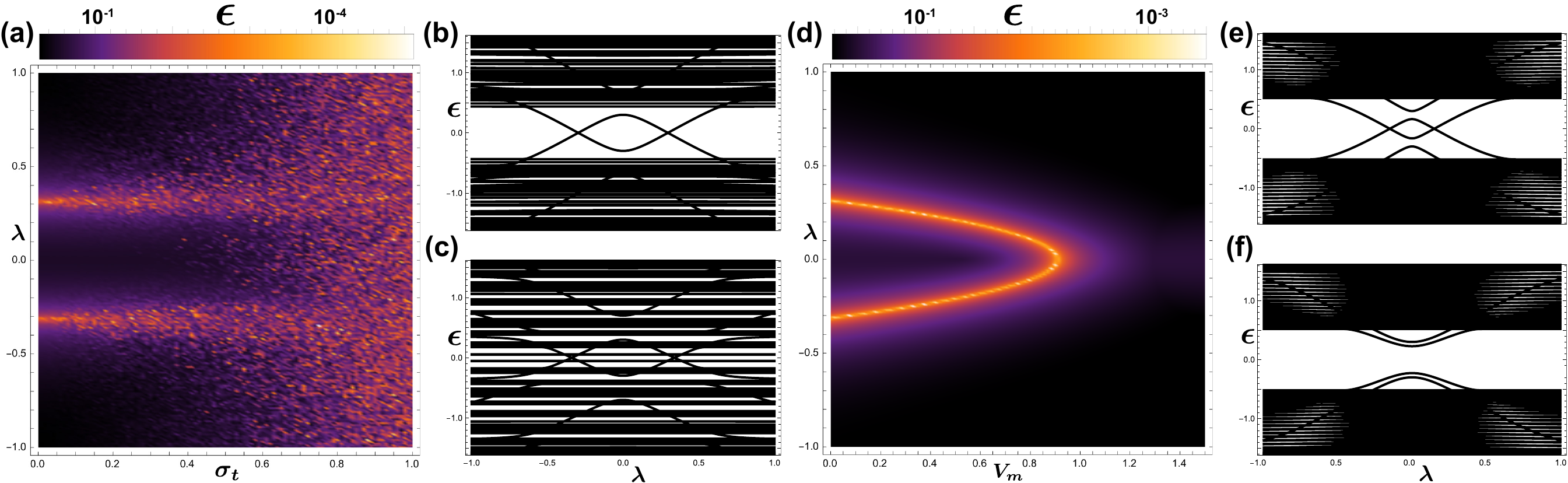}
    \caption{(a) The variation of lowest energy level (nonnegative) with $\lambda$ and disorder strength $\sigma_t$. Level crossings are obvious at weak disorder and become obscured when the disorder is strong enough. Comparing energy spectrum at $\sigma_t=0.3$ in (b) and $\sigma_t=0.8$ in (c), we find that disorder in $t_1$ term mainly influences the bulk gap. (d) The variation of lowest energy level (nonnegative) with $\lambda$ and impurity strength $V_m$. The impurity is added solely at site $j=1$. The level crossings move under the variation of impurity strength and disappear when boundary gap closes. Energy spectrum at two representative impurity strength, $V_m=0.7$ and $V_m=1.2$, are plotted in (e) and (f) respectively. In all the figures, $\bar t_1=1$, $t_2=2$, $m=0.5$, $\delta m=0.2$.}\label{figs3}
\end{figure*}

In this section, we derive the effective boundary Hamiltonian of the 2D model for an arbitrary edge. 

In the absence of $m$ and $\delta m$ terms, the bulk gap closes at $\Gamma$ point when $t_1=t_2$. Near this critical point, we write down the continuum model by expanding the bulk Hamiltonian at $\Gamma$ point up to second order in $\bm k$, which reads
\begin{equation}
    H(k_x,k_y) = [(t_2-t_1)-\frac{t_2}{2}(k_x^2+k_y^2)]\tau_y-t_2(k_x\tau_x+k_y\tau_z\sigma_z).
\end{equation}
To derive the effective Hamiltonian of an arbitrary edge, we consider another coordinate system $k_1k_2$ that is obtained by rotating $k_xk_y$ system counterclockwise by an angle $\phi$, as illustrated in Fig. \ref{figs2}. Coordinates in the two systems are related by
\begin{equation}
    k_x = k_1\cos\phi-k_2\sin\phi,\ k_y=k_1\sin\phi+k_2\cos\phi. \label{eq:k1k2}
\end{equation}
Substituting Eq.(\ref{eq:k1k2}) into the continuum Hamiltonian, we have 
\begin{align}
    &H'(k_1,k_2) = [(t_2-t_1)-\frac{t_2}{2}(k_1^2+k_2^2)]\tau_y \\
    &-t_2[(k_1\cos\phi-k_2\sin\phi)\tau_x+(k_1\sin\phi+k_2\cos\phi)\tau_z\sigma_z].\nonumber
\end{align}
We further rotate the inner basis with $U_\phi = e^{-i\phi/2\tau_y\sigma_z}$, and the resulting Hamiltonian for finite $m$ and $\delta m$ takes the form
\begin{align}
    \tilde H(k_1,k_2) & =  U H_1(k_1,k_2) U^\dagger \\
    & =[(t_2-t_1)-\frac{t_2}{2}(k_1^2+k_2^2)]\tau_y \nonumber\\
    & -t_2(k_1\tau_x+k_2\tau_z\sigma_z) \nonumber\\
    & -\delta m\tau_z\sigma_y - m\sigma_y e^{i(\phi-\theta)\tau_y\sigma_z}\nonumber
\end{align}
To derive boundary Hamiltonian, we consider a semi-infinite system with edges along $k_2$ direction (blue lines in Fig. \ref{figs2}). Hamiltonian for this semi-infinite 2D system is obtained by making a substitution $k_1\rightarrow -i\partial_x$ while keeping $k_2$ intact, which leads to
\begin{align}
    \tilde{H}(-i\partial_x,k_2) &= [(t_2-t_1)+\frac{t_2}{2}\partial_x^2]\tau_y+i t_2 \partial_x \tau_x \nonumber\\
    & -\frac{t_2}{2}k_2^2\tau_y-t_2 k_2\tau_z\sigma_z \nonumber\\
    & -\delta m\tau_z\sigma_y - m\sigma_y e^{i(\phi-\theta)\tau_y\sigma_z}\label{eq:hamhybrid}
\end{align}
In the absence of $m$ and $\delta m$ terms, the model supports two Majorana zero modes at $k_2=0$ in topological phase, which are localized on the edge. Eigenstates for these two modes are obtained by solving Schr$\ddot{\text{o}}$dinger equation for $m=\delta m=0$, i.e.,
\begin{equation}
    \tilde{H}(-i\partial_x,k_2=0)|\psi\rangle=0.
\end{equation}
The direction of an arbitrary edge is represented by a unit vector $\hat n_\phi$ pointing along its normal direction outwards, as shown in Fig. \ref{figs2}. In the following, we simply refer it to edge $\hat n_\phi$. Eigenstates of the two zero modes at edge $\hat n_\phi$ is given by 
\begin{align}
    &|\psi_n\rangle = c(e^{\eta_1 x}-e^{\eta_2 x})|\varphi_n\rangle,\ n=1,2 \\
    &|\varphi_1\rangle = (1\ 0)^T\otimes(1\ 0)^T,\ \ |\varphi_2\rangle = (0\ 1)^T\otimes(0\ 1)^T, \nonumber
\end{align}
where $\eta_{1/2}$ are the roots of $\frac{t_2}{2}\eta^2-t_2\eta+(t_2-t_1)=0$. We then obtain the effective edge Hamiltonian by projecting the bulk Hamiltonian in Eq.(\ref{eq:hamhybrid}) into eigenspace spanned by basis $\{|\psi_1\rangle,|\psi_2\rangle\}^T$, which reads
\begin{equation}
    H_E(\phi)=t_2 k_2 s_z+\delta m s_y - m\cos(\phi-\theta)s_y,
\end{equation}
with $s$ being Pauli matrices acting in the zero mode basis. From the effective edge Hamiltonian, we immediately obtain the edge gap $\Delta(\phi) = \delta m - m\cos(\phi-\theta)$, whose size as well as sign depends on edge orientation. Majorana corner states appear whenever gaps of adjacent edges take opposite signs.

\section{D. Effects of disorder and impurity on level crossings}

\begin{figure*}[t]
    \includegraphics[width=\textwidth]{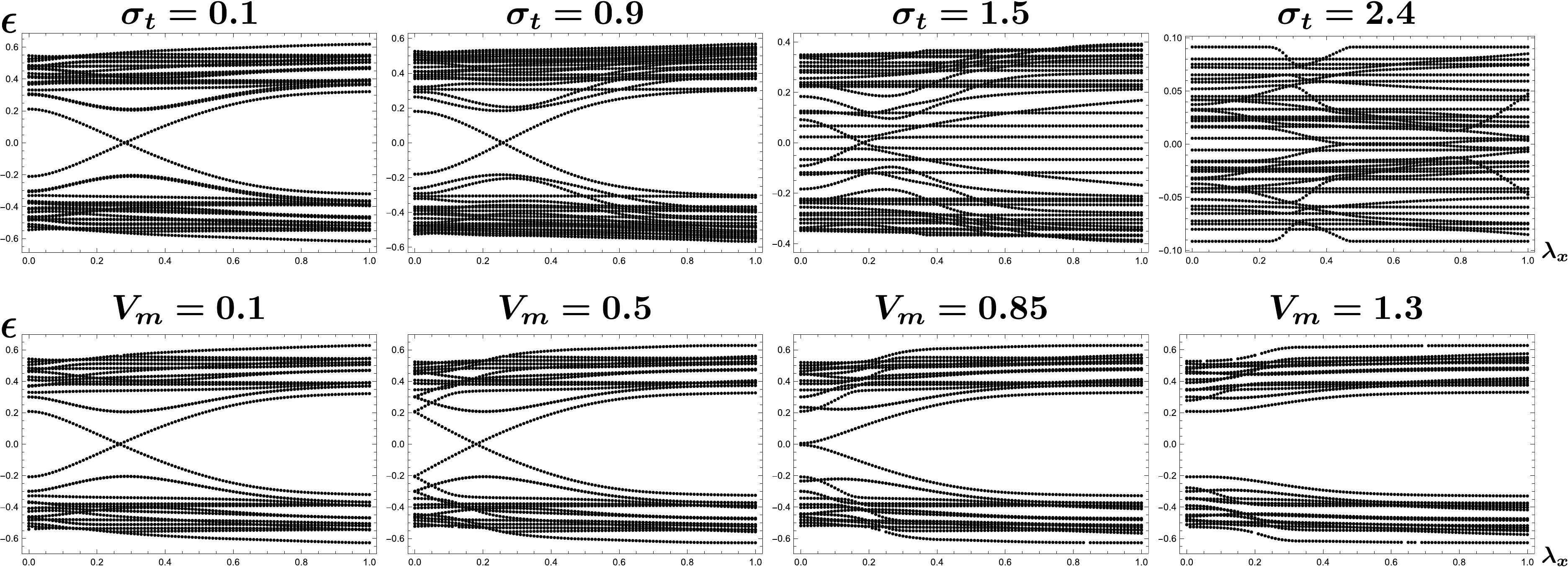}
    \caption{Fermi level crossings in a 2D system with disorder (upper panel) and boundary impurities (lower panel). Boundary impurities are added uniformly on the left edge. The system size is $40\times 40$, and $\bar t_1=1$, $t_2=2$, $m=0.7$, $\delta m=0.2$, $\theta=\pi/4$.}\label{figs4}
\end{figure*}

\begin{figure*}[t]
    \includegraphics[width=\textwidth]{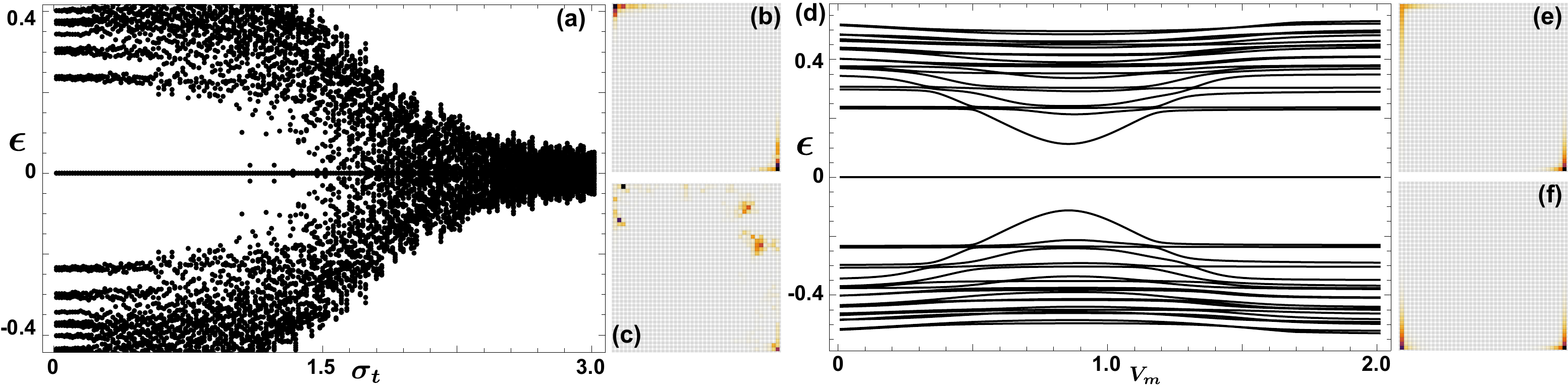}
    \caption{(a) Evolution of energy spectrum for a finite 2D system with bulk disorder $\sigma_t$. (b) and (c) Probability distributions of energy states closest to zero energy for $\sigma_t=0.9$ and $\sigma_t=2.4$.  Majorana corner states remain stable at weak disorder, and couple with bulk states when disorder is strong enough. (d) Evolution of energy spectrum with impurity strength $V_m$. The impurities are added uniformly on the left edge. Bulk gap is not influenced. Boundary gap of the left edge is expected to close and reopen with the variation of $V_m$. Due to finite size effect, the edge gap doesn't really close, but the topology of left edge may change when impurity strength is strong enough. Consequently, one of the Majorana corner states is transferred to neighboring corner, as can be seen by comparing the cases in (e) with $V_m=0.7$ and (f) with $V_m=1.2$. In all the figures, $\bar t_1=1$, $t_2=2$, $m=0.7$, $\delta m=0.2$, $\theta=\pi/4$.}\label{figs5}
\end{figure*}

In this section, we investigate the stability of level-crossing points against bulk disorder and boundary impurities, and study their influences on Majorana corner states in 2D.

As we pointed out in the main text, level crossings are protected by fermion parity conservation. A single crossing cannot disappear unless bulk or boundary gap is closed. Therefore, if the disorder or impurity doesn't close the two gaps, we can expect the level crossings to persist.

First, let us consider bulk disorder of $t_1$ term in the 1D model, which follows Gaussian distribution with mean value $\bar t_1$ and standard deviation $\sigma_t$. We plotted the lowest energy level (nonnegative) in $(\sigma_t,\lambda)$ parameter space, as shown in Fig. \ref{figs3}(a). For weak disorder (small $\sigma_t$), the level crossing points remain stable, as verified by energy spectrum shown in Fig. \ref{figs3}(b). With the increase of bulk disorder, more and more states move close to zero energy, indicating the closure of bulk gap, as Fig. \ref{figs3}(c) demonstrates. In the latter case, there is no longer any level crossing. Therefore, the disorder in $t_1$ term mainly influences the bulk gap and is expected to close the gap when it is strong enough.

In contrast to $t_1$ term, $m$ or $\delta m$ term would influence the boundary gap. We consider impurities of strength $V_m$ at boundary site $j=1$, which has the the same form as $m$ term, i.e., $V_m \Gamma_1^T \sigma_y \Gamma_1$. In Fig. \ref{figs3}(d), we find that with the increase of $V_m$, the two level-crossing points at $\pm \lambda$ move towards $\lambda=0$ and annihilate with each other where boundary gap closes. From Fig. \ref{figs3}(e) and (f), we find that the bulk gap doesn't change in this process, but the boundary gap closes and reopens, accompanied by the disappearance of level crossings.

Turning to 2D system, the topological invariant introduced in the main text is determined from level crossings at high symmetry momenta, and hence relies on translation symmetry. When disorder or impurities break translation symmetry, we can no longer say that the crossing appears at 1D subsystem with $K=0$ or $\pi$, but have to look at the 2D spectrum instead. We should emphasize that $\eta_a$ is the sum of the crossings at $K=0$ and $K=\pi$, and doesn't necessarily equal the number of crossings that appear while a toroidal system is deformed into a cylindrical one. This is because $K=\pi$ is not allowed when a periodic system has an odd number of unit cells but is allowed under anti-periodic boundary condition. When the numbers of unit cells along both directions are even, we could safely say that the two numbers are equal to each other. In Fig. \ref{figs4}, we show the energy spectrum of a $40\times 40$ lattice, and Fermi level crossings indeed survive under weak bulk disorder and boundary impurities that are added uniformly on the left edge.

Considering that the higher-order topology is intimately related to Fermi level crossings, we could expect Majorana corner states to be robust under these perturbations. In Fig. \ref{figs5}(a) and (d) we plotted the variations of energy spectrum at open boundaries with bulk disorder of $t_1$ term, as well as boundary impurities $V_m$ that are uniformly distributed on the left edge. Indeed, Majorana corner states survive under weak disorder, and disappear when disorder becomes so strong that the bulk gap closes, as shown in Fig. \ref{figs5}(b) and (c). Adding impurities on one edge only influences the boundary spectrum. The topology of this particular edge would change when the impurities are strong enough, so is the topological difference between it and neighboring edges. As a result, Majorana corner states do not disappear but may hop from one corner to an adjacent one, as shown in Fig. \ref{figs5}(e) and (f).

\begin{figure}[t]
    \includegraphics[width=0.48\textwidth]{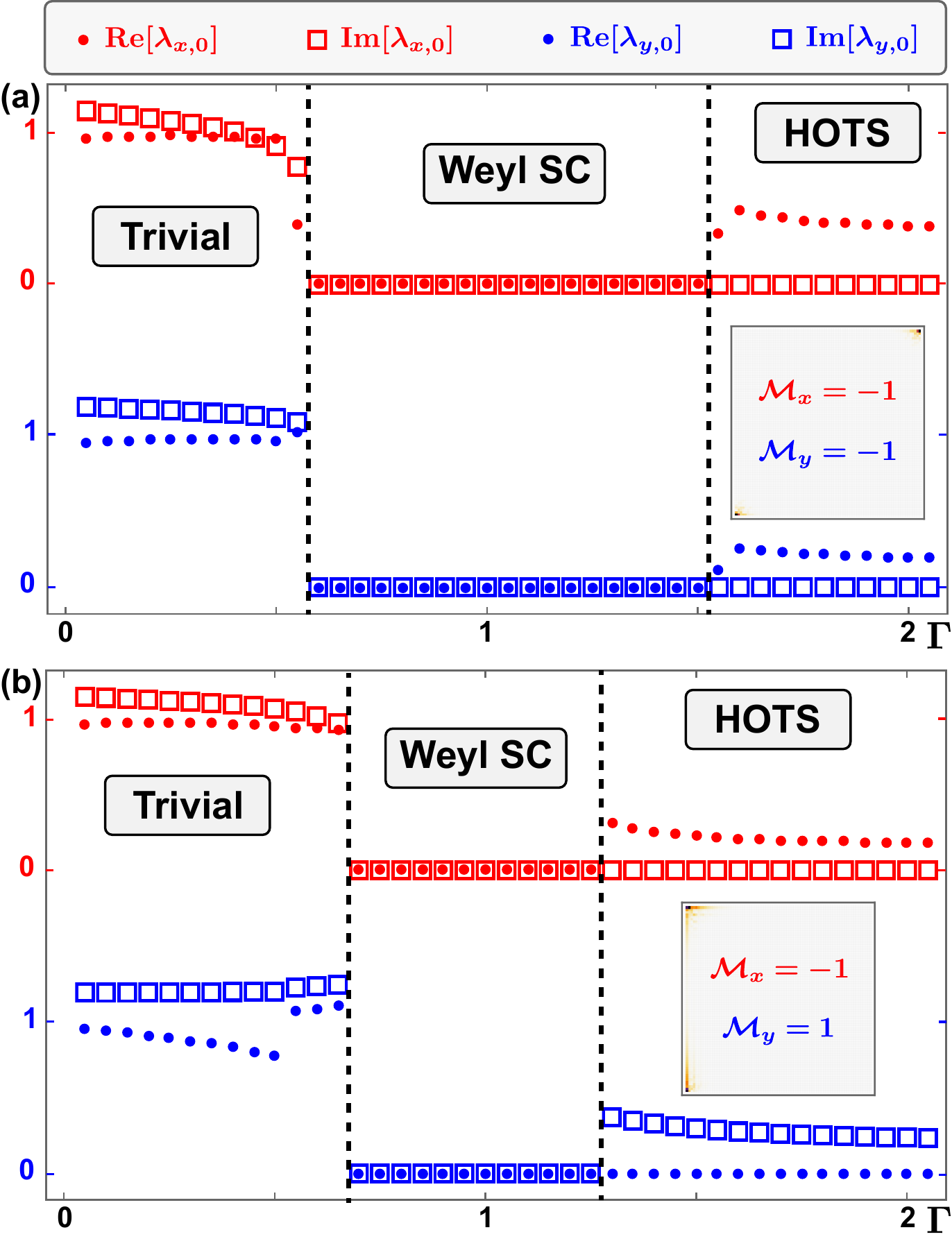}
    \caption{Fermi Level crossings in trivial, Weyl superconductor (SC) and higher-order topological superconductor (HOTS) phases. (a) $\phi=\pi$. Both Majorana numbers take $-1$ in higher-order phase, suggesting the two Majorana corner states (shown in the inset) sit at opposite corners.  (b) $\phi=3\pi/4$. Only $\mathcal M_x=-1$. The two Majorana corner states sit at adjacent corners. Note that level-crossings only occur at $K=0$ with given parameters. In all the figures, $t=\alpha=\Delta=2\Delta_Z=1$, $\mu=0$, $\theta=\pi/3$.}\label{figs6}
\end{figure}

\section{E. Rashba Bilayer}

In this section, we apply the topological invariant proposed in the main text to a Rashba bilayer superconducting system that is known to support higher-order phases \cite{volpez2019}. There are three key ingredients that make it a higher-order superconductor: Rashba spin-orbit coupling, in-plane Zeeman field and phase difference of $s$-wave pairing between the two layers. The model Hamiltonian in $k$-space is given by
\begin{align}
    H(\bm k) = &[\mu+4t-2t(\cos k_x+\cos k_y)]\tau_z\\
    + & \alpha(\sin k_y s_x-\sin k_x\tau_z s_y)\sigma_z \nonumber\\
    + & \frac{\Delta}{2}\tau_y s_y[(\sigma_0+\sigma_z)+e^{i\phi \tau_z}(\sigma_0-\sigma_z)]\nonumber\\
    + & \Gamma \tau_z\sigma_x + \Delta_Z(\cos \theta \tau_z s_x+\sin\theta s_y),\nonumber
\end{align}
where $\tau$, $s$ and $\sigma$ are Pauli matrices that act in Nambu, spin and layer space respectively. In this model, $\mu$ represents chemical potential, $t$ is amplitude of nearest-neighboring hopping, $\alpha$ represent the strength of Rashba spin-orbit coupling, $\Delta$ is the pairing of upper layer, $\phi$ is phase difference between the two layers, $\Gamma$ denotes the coupling between two layers, and $\Delta_Z$ is in-plane Zeeman field with $\theta$ being its direction.

For $\phi=\pi$ and $\Delta_Z<\Delta$, the model realizes higher-order phase with two Majorana corner states at opposite corner when $\Gamma>|\Delta+\Delta_Z|$ and is in trivial phase when $\Gamma<|\Delta-\Delta_Z|$. In between the two phases, the system becomes a Weyl superconductor. Indeed, Fermi level crossings (real root $\lambda\in (0,1)$) appear in the higher-order phase, as can be seen in Fig. \ref{figs6}(a). In the Weyl superconductor phase, level crossings appear exactly at $\lambda=0$. It should be noted that the higher-order phase persists for a wide range of $\phi$, in which case Majorana corner states may reside at neighboring corners instead of opposite corners, as shown in the inset of Fig. \ref{figs6}(b).

\end{document}